\documentclass[12pt]{iopart}
\usepackage{iopams} 
\usepackage[latin1]{inputenc}
\usepackage{graphicx}
\usepackage{url}

\begin{document} 
 
\title[PT symmetry -- graphene  -- surface plasmon]{Lasing condition for trapped modes in subwavelength--wired PT--symmetric resonators}
\author{Mauro Cuevas} 
\address{Consejo Nacional de
Investigaciones Cient\'ificas y T\'ecnicas (CONICET) and Universidad Austral, Facultad de Ingenier\'ia, Pilar
Mariano Acosta 1611 B1629WWA-Pilar-Buenos Aires
Argentina}

\author{Mojtaba Karimi} 
\address{ Faculty of Physics, University of Tabriz, 51664, Tabriz, Iran}

\author{Carlos J. Zapata--Rodr\'iguez} 
\address{Department of Optics and Optometry and Vision Science, University of Valencia, Dr. Moliner 50, Burjassot 46100, Spain}

\ead{mojtaba.karimihabil@gmail.com}

\begin{abstract} 
The ability to control the laser modes within a subwavelength resonator is of key relevance in modern optoelectronics. 
This work deals with the theoretical research on optical properties of a PT--symmetric nano--scaled  dimer  formed by two dielectric wires, one is with loss and the other with gain, wrapped with graphene sheets. We show the existence of two non--radiating trapped modes which transform into radiating modes by increasing the gain--loss parameter. Moreover, these modes reach the lasing condition for suitable values of this parameter, a fact that makes these modes to achieve an ultra high quality factor that is manifested on the response of the structure when it is excited by a plane wave. Unlike other mechanism that transform trapped modes into radiating modes, we show that the  variation of gain--loss parameter in the balanced loss--gain structure here studied  leads to a variation in the phase difference between induced dipole moments on each wires, without appreciable variation in the modulus of these dipole moments. We provide an  approximated method that reproduces the main results provided by the rigorous calculation. Our theoretical findings reveal the possibility to develop unconventional optical devices and structures with enhanced functionality.
\end{abstract} 



\maketitle

\section{Introduction} 


It is known that, under certain conditions, a plasmonic system can support trapped electromagnetics modes, 
which are electromagnetic non radiating oscillations that stay localized inside the structure \cite{Zoudi}. These plasmon modes are  
characterized by an ultra--high quality factor provided that the structural symmetry remains unbroken. 
In fact, the narrow resonances characterizing these mode excitations rely on the breaking degree of the  symmetry which allows the trapped mode to couples with free photons \cite{Fedotov}.  In this sense, plasmon trapped modes can be considered as 
 a kind of symmetry protected  states. 
Due to Ohmic loss in the plasmonic material, the eigenfrequencies of the trapped states are complex valued. However, by introducing gain material elements into the structure its possible to compensate this  material loss and, as a  consequense, to reach the lasing condition for which the eigenfrequency associated to an eigenmode is real valued \cite{Zheludev}. 

Because of their fundamental properties as well as their  potential applications, the study of hybrid systems composed of gain material media and plasmonic materials, which are loss media, is a topic of continuous increasing interest \cite{spacer1,spacer2,Alu,spacer3,spacer4}.  In areas such as condensed matter and surface optics, the amplification of eigenmodes by stimulated emission of radiation has played a key role in the interpretation of a wide variety of experiments, the understanding of various fundamental properties of solids and  the engineering of nanolaser devices   \cite{Oultonnature,MocciaIEEE,Luis}. In particular, new phenomena associated with parity--time (PT) symmetry have been observed in optical systems with balanced loss and gain \cite{MiriOL,AluPRX}. These optical systems belong to a large family of  non--Hermitian systems, which can have a real spectrum provided that the system be invariant under combined operations of parity (P) and time--reversal (T) symmetry \cite{Bender,Ruter}.

Possibilities have been widened towards PT symmetric structures that incorporate graphene as plasmonic material. Doped  graphene allows the propagation of surface plasmons with low Ohmic losses, \textit{i.e.}, with high quality factor, from terahertz to near--infrared range \cite{depineguias}. Moreover, the plasmon resonance spectrum, and consequently the optical responses of the system, can be tuned by varying the doped level on graphene. In this context, several works have focused on the influence that long living and tunable surface plasmons on graphene have on the optical responses of a PT--symmetric system \cite{ChenPRApplied,LinOL,Zhernovnykova,Zhang}. 

The present work contains all the ingredients above entered and focusing on the study of a sub--wavelength  graphene plasmonic system with balanced volume losses. 
In particular, our system is composed of two parallel dielectric cylinders, one is with loss and the other with the same level of gain, both wrapped with graphene shets. 
Here, the sole effect of graphene is to provide LSP resonances in sub--wavelength wires. Although other more complex structures such as an array of many or infinite cylinders can be designed to present PT--symmetry, our motivation is based in the   simplicity of the dimer structure that we have chosen, since it allows us to understand the underlying physics behind the PT--symmetry.  We use a rigorous formalism based on Mie theory to calculate the eigenmodes dependence with the gain--loss parameter for both trapped and radiating eigenmodes.  
To do this, we solve the boundary value problem with the corresponding boundary conditions without an incident wave (homogeneous problem).   The procedure requires the analytic continuation of the eigenvalue, the frequency in our case, in the complex plane.  
Analytic continuation is inevitable, even in the case of media
with no intrinsic losses, since the open nature of our resonator generates  non--null imaginary
parts due to radiation losses. 

In addition, as it is well established nowadays,   
the modal lasing analysis in an open plasmonic resonator, as our  PT--symmetric structure, can be carried out by 
 applying the lasing eigenvalue problem (see \cite{nosich} and references therein),  
in which the modal  eigenfrequencies are assumed to be real valued at  the lasing condition (or lasing threshold). Following this concept, we solve the homogeneous  problem to find the  real valued eigenfrequencies, 
gain--loss parameters and 
chemical potentials for each modal lasing condition. Moreover, we study the eigenmode influence on the optical response of the system when it is excited by a plane wave near the lasing condition.

By using the quasistatic approximation valid in the long wavelength limit, 
we show that
 despite the graphene ohmic losses, a fact that gives rise to a complex spectrum, the structure exhibits a set of properties in common with a PT--symmetric system. For instance, two eigenmode branches coalesce at an exceptional point and, for the gain--loss parameter above certain threshold, these both branches are repelled in the direction of the imaginary part of frequency, making that one of these branches achieves the lasing condition.
In addition, we demonstrate that branches eigenmode, which are trapped modes in case of null gain--loss parameter, transform into radiating modes that can be excited by plane wave incidence when the gain--loss parameter is increased. Unlike previous works \cite{Tuz,CSZ}, where  transitions from trapped to radiating eigenmodes are achieved by producing a difference between the modulus of individual dipole moments on each particle forming the dimer, \textit{i.e.}, by producing a weakly asymmetry with respect to the center of the dimer modifying the dipole moment amplitudes, here, we show that the increment of the gain--loss parameter leads to a change in the phase difference between these individual dipole moments, maintaining their modulus values constant. 

This paper is organized as follows. In section \ref{teoria} we present a brief description of the rigorous method used in this work to  calculate the scattering of a dimmer composed of  two graphene wires. From this method, using the quasistatic approximation, we deduce analytical  expressions for eigenfrequencies and eigenvectors as a function of geometrical and constitutive parameters that explain the main features calculated with the rigorous method. In section \ref{resultados} we present results   of two parallel dielectric 
cylinders tightly coated with a graphene layer, one of then with small inner losses and the other one with the same level of gain. 
 Concluding remarks are provided in Section \ref{conclusiones}. The Gaussian system of units is used and an $\mbox{exp}(-i\, \omega\, t)$ time--dependence is implicit throughout the paper, where $\omega$ is the angular frequency, $t$ is the time coordinate, and $i=\sqrt{-1}$. The symbols Re and Im are used for denoting the real and imaginary parts of a complex quantity, respectively.

\section{Theory} \label{teoria}

\subsection{Rigorous description of the fields and scattering efficiencies}

\begin{figure}
\centering
\resizebox{0.60\textwidth}{!}
{\includegraphics{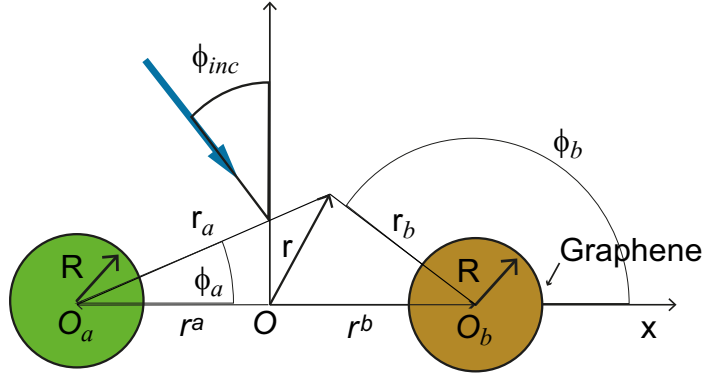}}
\caption{\label{fig:epsart} Schematic illustration of the system composed by $2$ dielectric cylinders wrapped with graphene sheets.
}\label{sistema}
\end{figure}

We consider a cluster  consisting of two parallel and non overlapping cylindrical dielectric wires, one with gain, $\varepsilon_a=\varepsilon_1-i\varepsilon_i$ ($\varepsilon_i>0$),  and the other with equal loss, $\varepsilon_b=\varepsilon_1+i\varepsilon_i$, as shown in Figure \ref{sistema}. Both wires have the same radius $R_a=R_b=R$ and are wrapped with a graphene sheet. The system is embedded in a lossless and non--magnetic dielectric denoted as medium $v$ with permittivity $\varepsilon_v$.
 In this case, a PT symmetry around the central axis, denoted by $O$, is fulfilled. We assume that the radius $R$ is sufficiently large to describe the optical properties of the wires as characterized by the same surface conductivity as planar graphene (see appendix \ref{grafeno}). We denote by $r_j(\mathbf{r}),\,\phi_j(\mathbf{r})$ ($j=a,\,b$) the polar  coordinates of a point at position $\mathbf{r}$ with respect to the local origin $O_j$. A plane wave radiation  impinges on the wires with an angle of incidence $\phi_{inc}$ with respect to the $y$ axis.  Although some enhanced optical effects related to invisibility modes are observed for $s$ polarization (electric field along the $z$ axis) \cite{Carlossci}, this work focus on $p$ polarization 
 (magnetic fields along the $z$ axis) for which 
  the electric field  in the graphene coating induces electric currents directed along the azimuthal direction $\phi_j(\mathbf{r})$ and LSPs exist in the graphene circular cylinder. 
In this way,   
the incident magnetic field (along the $z$ axis) can be written in a system linked to the $j$--cylinder as \cite{CSZ}, 
\begin{equation}\label{campo_incidente}
H_{inc}(\mathbf{r}) = e^{i k_v r^j \sin(\phi_{inc}-\phi^j)} \sum_{m=-\infty}^{+\infty} (-1)^m J_m(k_v r_j(\mathbf{r})) e^{i m \phi_l(\mathbf{r})} e^{-i m \phi_{inc}}
\end{equation}
where $r^j,\,\phi^j$ are the polar coordinates of the $j$--cylinder, $k_v=\sqrt{\varepsilon_v} \frac{\omega}{c}$,  is the modulus of the photon wave vector in medium $v$, $\omega$ is the angular frequency, $c$ is the vacuum speed of light and $J_m(x)$ is  the $n$th Bessel function.  The scattered magnetic field in medium $v$ ($r_j(\mathbf{r})>R$) can be written as a superposition of the field scattered by each of the cylinders, 
\begin{equation}\label{campo_scattereado}
H^{(v)}_s(\mathbf{r}) = \sum_{j=a,b} \sum_{m=-\infty}^{+\infty} b_{j\,m} H_m(k_v r_j(\mathbf{r})) e^{i m \phi_j(\mathbf{r})}
\end{equation}
where $H_m(x)$ is the $n$th Hankel functions of the first kind. Note that the $j$th term of the summation corresponds to the field scattered by the $j$th cylinder linked to the local system with origin $O_j$. In the region inside the cylinders, $r_j(\mathbf{r})<R$, the transmitted field is written as 
\begin{equation}\label{campo_transmitido}
H^{(j)}(\mathbf{r}) = \sum_{m=-\infty}^{+\infty} a_{j\,m} J_m(k_j r_j(\mathbf{r})) e^{i m \phi_j(\mathbf{r})}
\end{equation}
where $j=a,\,b$. To find the unknown complex amplitudes of the reflected $b_{j m}$ and transmitted $a_{j m}$ fields (\ref{campo_scattereado}) and (\ref{campo_transmitido}), we use  the usual boundary  conditions   and the addition theorem for Bessel and Hankel functions \cite{abramowitz}. This theorem allow us to write one of the terms in (\ref{campo_scattereado}), associated to the scattered field of one of the cylinders (for example, the  $j=b$ cylinder) in the other local coordinates (the $j=a$ cylinder). In this way, the scattered field (\ref{campo_scattereado}) will be represented in the form of expansions in Hankel functions 
written in the local coordinate $j=a$. By replacing this expression and Eq. (\ref{campo_transmitido}) with $j=a$ into the boundary conditions along the surface $r_a=R$ of the $j=a$ cylinder, one obtain a set of $2$ equations for the   $2 \times 2$  unknown  amplitudes. Similarly, we can write the scattered field (\ref{campo_scattereado}) in the local coordinate $j=b$ and use the boundary conditions on the surface of the $j=b$ cylinder to obtain other set of 2 equations for the   $2 \times 2$  unknown  amplitudes.  However, we closely follow a variant of the method, developed in \cite{Maystre}, that allows  to reduce to half the dimension of the system of equations. The detailed of this implementation has been given in \cite{CSZ}, and leads to the following system of equations for the amplitudes $b_{j m}$ 
\begin{equation}\label{sistema_ecuaciones_matricial}
\left[
\begin{array}{llll}
\,\,\,\,\,\,\,\,\,\,\,\,\overline{\overline{\mathbf{I}}} & -\overline{\overline{\mathbf{S}}}_a\cdot\overline{\overline{\mathbf{T}}}_{a\,b}  \\
-\overline{\overline{\mathbf{S}}}_b\cdot \overline{\overline{\mathbf{T}}}_{b\,a} & \,\,\,\,\,\,\,\,\,\,\,\,\overline{\overline{\mathbf{I}}} \\
\end{array} \right] 
\left( 
\begin{array}{lll}
\overline{\mathbf{b}}_a\\
\overline{\mathbf{b}}_b\\
\end{array} \right)=
\left( 
\begin{array}{lll}
\overline{\overline{\mathbf{S}}}_a\cdot \overline{Q}_a\\
\overline{\overline{\mathbf{S}}}_b\cdot \overline{Q}_b\\
\end{array} \right),
\end{equation}
where $\overline{\mathbf{b}}_l$ and $\overline{\mathbf{Q}}_l$ are  vectors whose coordinates are the elements $b_{l m}$ and 
\begin{eqnarray}\label{Q}
Q_{j m}= e^{i k_v r^j \sin(\phi_{inc}-\phi^j)} (-1)^m e^{-i m \phi_{inc}},
\end{eqnarray}
respectively, $\overline{\overline{\mathbf{T}}}_{lj}$ is the matrix with elements
\begin{eqnarray}\label{T}
 T_{l j n m}= H_{n-m}(k_v r_l^j) e^{i (n-m) \phi_l^j}, 
\end{eqnarray}
and 
$\mathbf{S_j}$ is the matrix with elements $S_{j m n}=s_{j m} \,\delta_{m n}$, where $s_{j m}$ are the elements  of the scattering matrix associated to the $j$th cylinder \cite{cuevasJQSRT214} 
\begin{eqnarray}\label{S}
s_{j\,m} = 
& \frac{[k_j \varepsilon_v J_m(y) J^{'}_m(x)-k_v \varepsilon_j J_m^{'}(y) J_m(x)-\frac{4\pi \sigma}{c k_0} i k_j k_v J^{'}_m(x) J^{'}_m(y)]}{k_v\varepsilon_j J_m(x) H^{'}_m(y)-k_j \varepsilon_v J^{'}(x)H_m(y)+\frac{4\pi \sigma}{c k_0} i k_j k_v J^{'}_m(x) H^{'}_m(y)}
\end{eqnarray}
where $x=k_j R$, $y=k_v R$ and the prime denotes the derivative with respect to the argument.

Knowing the total electromagnetic field allows us to calculate the scattering cross sections. The time--averaged scattered power is calculated  from the integral of the radial component of the complex Poynting vector flux through an imaginary cylinder of length $L$  of radius $\rho_0$ which envelops the graphene wire system (see Fig. 1),
\begin{equation}\label{potencia}
P_s=\rho_0 L \int_0^{2 \pi} \left\langle \mathbf{S}\right\rangle \cdot \hat{r}\, d\phi,
\end{equation}
\begin{eqnarray}\label{poynting}
\left\langle \mathbf{S}(\rho_0,\phi)\right\rangle \cdot \hat{r}=\frac{c}{8 \pi} \Re \left\{ E_{s, \phi} H_s^{*}\right\}= 
\frac{c}{8 \pi} \Re \left\{\frac{1}{i k_0 \varepsilon_v} \frac{\partial \, H_s}{\partial r} H_s^{*}\right\}.
\end{eqnarray}
It is convenient to express each of the terms in Eq. (\ref{campo_scattereado}) in a same coordinate $O$ \cite{CSZ}.  Substituting the obtained expression into Eq. (\ref{poynting}) we obtain
\begin{eqnarray}\label{poynting2}
\left\langle \mathbf{S}(\rho_0,\phi)\right\rangle \cdot \hat{r}=
\frac{c}{8 \pi} \Re \left\{\frac{1}{i \sqrt{\varepsilon_v}}  \sum_{m,n} B_m B_n^{*} H'_m(k_v \rho_0) H^{*}_m(k_v \rho_0) e^{i(m-n)\phi}\right\},
\end{eqnarray}
where 
\begin{eqnarray}\label{Bm}
B_m=\sum_{j=a,b} \sum_{q=-\infty}^{+\infty} b_{j m} J_{m-q}(k_v r^j) e^{i(q-m) \phi^j}.
\end{eqnarray}
Inserting Eq. (\ref{poynting2}) into (\ref{potencia}) and taking into account the wronskian $W\left\{J_m(x),Y_m(x)\right\}=2/(\pi x)$ ($x=k_v \rho_0$),  after some algebraic manipulations, we obtain the power scattered by the cylinders,  
\begin{equation}\label{potencia2}
P_s=\frac{c^2 L}{2\pi \omega \varepsilon_v} \sum_{m=-\infty}^{+\infty} |B_m|^2.
\end{equation}
The scattering cross section is defined as the ratio between the total power scattered by the cylinders, given
by Eq. (\ref{potencia2}), and the incident power $P_{inc}$ intersected by the area of all the cylinders. 

It is known that the scattering cross section and the near to the cylinders field are strongly affected by complex singularities in the field amplitudes $b_{j m}$. Singularities occur at complex locations and they represent  the frequency of the eigenmodes supported by the cylinders system.  
Complex frequencies of these modes are obtained by solving the homogeneous problem, \textit{i.e.}, by imposing the vectors $\overline{\mathbf{Q}}_a$ and $\overline{\mathbf{Q}}_b$ in Eq. (\ref{sistema_ecuaciones_matricial}) be zero. Then, a condition to determine eigenfrequencies is to  require  the determinant of the matrix  in Eq. (\ref{sistema_ecuaciones_matricial}) to be zero,
\begin{equation}\label{dispersion}
\mbox{det}\left[
\begin{array}{llll}
\,\,\,\,\,\,\,\,\,\,\,\,\overline{\overline{\mathbf{I}}} & -\overline{\overline{\mathbf{S}}}_a\cdot\overline{\overline{\mathbf{T}}}_{a\,b}\\
-\overline{\overline{\mathbf{S}}}_b\cdot\overline{\overline{\mathbf{T}}}_{b\,a} & \,\,\,\,\,\,\,\,\,\,\,\,\overline{\overline{\mathbf{I}}} \\
\end{array} \right]=0.
\end{equation} 
This condition corresponds to the full retarded  dispersion equation (FR) for  eigenmodes and it determines the complex frequencies $\omega=\omega_R+i\omega_I$ ($\omega_I<0$) in terms of all the geometrical and constitutive parameters of the system.

\subsection{Quasistatic approximation. A simple model based on two coupled electric dipoles} \label{QA}

 Although the  rigorous treatment represented by Eq. (\ref{dispersion}) 
gives us all the kinematics and dynamics characteristics of the PT--symmetric eigenmodes, the method lacks of analytical expressions that explain the main dependencies with both geometrical and constitutive parameters. For the purpose of showing this dependencies, by applying the quasistatic method, here,  we reduce the full treatment provided by the homogeneous part of Eq. (\ref{sistema_ecuaciones_matricial}) 
to a simple 2x2 matrix description as follows.

Assuming that the radius $R$ of cylinders is much smaller than the wavelength $\lambda=2\pi/\omega$, the problem can be treated using the dipole approximation where the dimer eigenfunctions calculated with (\ref{sistema_ecuaciones_matricial}) are,  as a good approximation, a superposition of single plasmons with angular momentum $m=\pm 1$ linked to each cylinder. In this way, only four coordinates, $b_{j\pm1}$ ($j=a,\,b$), define the amplitudes vector.

We first consider the case in which the induced dipole moments are along the $\pm x$ directions, \textit{i.e.}, the local magnetic field associated to each of the  cylinders has a dependence $\approx \sin \phi_j$ ($j=a,\,b$). As a consequence, the amplitudes $b_{j 1}=b_{j -1}$ ($j=a,\,b$). Here, the subscript $j -1$ stand for cylinder $j$ and angular momentum $m=-1$. 
In this way, 
the matrix equation for the modal amplitudes 
reduces to a 2$\times$2 matrix system for amplitudes $b_{a 1}$ and $b_{b 1}$, 
\begin{equation}\label{sistema_ecuaciones_matricial_2}
\left[
\begin{array}{llll}
1 & - 2\,s_a\,H_1(z)/z  \\
- 2\,s_b\,H_1(z)/z & 1 \\
\end{array} \right] 
\left( 
\begin{array}{lll}
b_{a 1}\\
b_{b 1}\\
\end{array} \right)=
\left( 
\begin{array}{lll}
0\\
0\\
\end{array} \right),
\end{equation}
where we have used $2 H_1(z)/z=H_0(z)+H_2(z)$, $z=k_v R_{ab}$ ($R_{ab}$ is the distance between cylinder centers). Using the small argument asymptotic expansions for Bessel and Hankel functions \cite{abramowitz}, it follows that $s_{j}$ can be written as
\begin{eqnarray}\label{s1}
s_j=\frac{\pi\, k_v^2}{4} i \alpha_j,
\end{eqnarray}
$j=a,\,b$, where
\begin{eqnarray}\label{alpha_1}
\alpha_j= R^2 \frac{\varepsilon_1\pm i\varepsilon_i-\varepsilon_v+g(\omega)}{\varepsilon_1\pm i\varepsilon_i+\varepsilon_v+g(\omega)}, 
\end{eqnarray}
are the dipolar polarizability of the cylinders $j=a$ or $j=b$, respectively, $g(\omega)=-\frac{\omega_g^2}{\omega^2+i \omega \gamma_g}$, $\omega_g^2=\frac{4 e^2 \mu}{ \hbar^2 R}$ is the effective plasma frequency for the dipolar mode \cite{CRD}. Note that the value of $\alpha_a$ differs from $\alpha_b$ in the sign of the gain--loss parameter $\varepsilon_i$, signs $+$ and $-$ correspond to $j=a$ and $j=b$ respectively. Near the resonance frequency, the polarizability $\alpha_j$ ($j=a,\,b$) can be written as
\begin{eqnarray}\label{alpha_1_approx}
\alpha_j= R^2 \frac{\varepsilon_1\pm i\varepsilon_i-\varepsilon_v+g(\omega_j)}{g'(\omega_j) (\omega-\omega_j)}=\frac{A_j}{\omega-\omega_j}, 
\end{eqnarray}
where $\omega_j=\omega_{jR}+i\omega_{jI}$ ($\omega_{jI}<0$) is the complex pole of $\alpha_j$, \textit{i.e.}, the eigenfrequency of the single graphene cylinder $j$, $g'(\omega)$ is the derivative of $g(\omega)$ and 
\begin{eqnarray}\label{A}
A_j=R^2\frac{\varepsilon_1\pm i\varepsilon_i-\varepsilon_v+g(\omega_a)}{g'(\omega_a)}.
\end{eqnarray}
%
The single cylinder eigenfrequencies for cylinders $a$ and $b$ are written as\cite{CRD}
\begin{eqnarray}\label{omega_1}
\omega_{j}= \frac{\omega_g}{\sqrt{\varepsilon_j+\varepsilon_v}}-i\frac{\gamma_g}{2}\approx \omega_0+i\omega_{jI}, 
\end{eqnarray}
where
\begin{eqnarray}\label{omega_R}
\omega_{0}= \frac{\omega_g}{\sqrt{\varepsilon_1+\varepsilon_v}},
\end{eqnarray}
\begin{eqnarray}\label{omega_I}
\omega_{jI}=-(\Gamma_{sup}\pm\Gamma_{core}), 
\end{eqnarray}
the signs $+$ and $-$ corresponds to $j=a$ and $j=b$, respectively, and 
\begin{eqnarray}\label{Gammas}
\Gamma_{sup}=\frac{\gamma_g}{2},\nonumber\\
\Gamma_{core}=\frac{1}{2}\varepsilon_i \frac{\omega_g}{(\varepsilon_1+\varepsilon_v)^{3/2}},
\end{eqnarray}
are the damping rates corresponding to the ohmic loss in graphene covers and dielectric cores, respectively. We have used the fact that $\varepsilon_i<<\varepsilon_1+\varepsilon_v$ in the last equality in Eqs. (\ref{omega_1}). 
From Eq. (\ref{omega_R})  we see that the real part of the eigenfrequencies 
do not depends on $\varepsilon_i$, 
a fact that also results by solving the fully retarded dispersion equation for a single graphene cylinder \cite{Leila}.

By taking into account the small argument $z=k_v R_{ab}$ in the Hankel functions, the off diagonal elements in the matrix  in Eq. (\ref{sistema_ecuaciones_matricial_2}) are written as
\begin{eqnarray}\label{elementos_no_diagonales}
-2\frac{\,H(k_v R_{ab})}{k_v R_{ab}} s_j= -2\frac{-2 i}{\pi (k_v R_{ab})^2} \frac{\pi k_v^2 i}{4} \alpha_j= -\frac{\alpha_j}{R_{ab}^2}.
\end{eqnarray}
Therefore, the matrix equation (\ref{sistema_ecuaciones_matricial_2}) takes the form 
\begin{equation}\label{sistema_ecuaciones_matricial_3}
\left[
\begin{array}{llll}
\omega-\omega_a &  -A_a/R_{ab}^2  \\
- A_b/R_{ab}^2 & \omega-\omega_b \\
\end{array} \right] 
\left( 
\begin{array}{lll}
b_{a 1}\\
b_{b 1}\\
\end{array} \right)=
\left( 
\begin{array}{lll}
0\\
0\\
\end{array} \right),
\end{equation}
where $A_j$ $j=a,\,b$ are given by Eq. (\ref{A}). Eq. (\ref{sistema_ecuaciones_matricial_3}) gives us a simple description for the dimer dynamic. For large separations, $A_j/R^2_{ab}<<\omega_a$  (or $R/R_{ab}<<1$), the matrix (\ref{sistema_ecuaciones_matricial_3}) is diagonal and  thus the eigenfrequencies correspond  to that of each individual graphene wires composing the dimer, \textit{i.e.}, the plasmonic wires do not  interact  between them. Taking into account the PT parameters, the real parts of both frequencies are the same whereas their imaginary parts differ in $2\Gamma_{core}$.    For small enough values of $R_{ab}$, extra diagonal terms take appreciable values and a splitting between  the real parts of eigenfrequencies occur. From Eq. (\ref{lado_derecho}) we can  see that this splitting is proportional to $R^2/R^2_{ab}\omega_0$.     
For system (\ref{sistema_ecuaciones_matricial_3}) to have a non--trivial solution, its determinant must be equal to zero, a condition which can be written as
\begin{eqnarray}\label{dispersion_2}
(\omega-\omega_a)(\omega-\omega_b)=\frac{A_a A_b}{R_{ab}^4}.
\end{eqnarray}
A detailed developed of the right hand side of this equation can be seen in appendix \ref{desarrollo}. By replacing  expression (\ref{lado_derecho}) into Eq. (\ref{dispersion_2}) and solving for $\omega$ eigenfrequencies, 
\begin{eqnarray} \label{omega}
\omega_{\pm}=-i\frac{\gamma_g}{2}+\omega_0 \pm \sqrt{-\frac{\gamma^2}{4}+\omega_{aI}\omega_{bI}+\frac{A_a A_b}{R_{ab}^4}}=\nonumber\\
\omega_0-i\frac{\gamma_g}{2} \pm 
\frac{\omega_0}{(\varepsilon_1+\varepsilon_v)}\sqrt{\frac{-\varepsilon_i^2}{4}+\frac{R^4}{R_{ab}^4}  \varepsilon_v^2 }
\end{eqnarray}
where in the last equality we have used Eq. (\ref{omega_I}).
\begin{figure}[htbp!]
\centering
\resizebox{0.95\textwidth}{!}
{\includegraphics{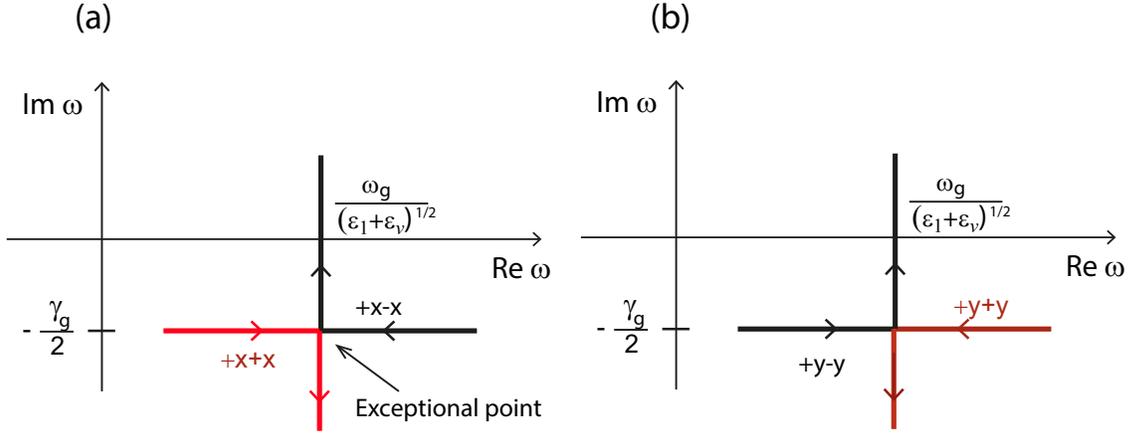}}
\caption{\label{fig:epsart} Trajectory in the complex plane of the eigenfrequencies (\ref{omega}) as a parametric function of the gain--loss parameter $\varepsilon_i$. Two branches,  (a) $+x$$-x$ and $+x$$+x$ and  (b) $+y$$-y$ and $+y$$+y$, are observed. As $\varepsilon_i$ increases,  the two eigenfrequencies approach each other until they coalesce in an exceptional point. After passing the exceptional point, they are repelled in their imaginary parts. 
 The real parts of the permittivity of the cylinders are $\Re \, \varepsilon_1=\Re \, \varepsilon_2=2.13$, $\varepsilon_v=1$, the radius $R_a=R_b=0.03\mu$m and the gap $\Delta=0.04\mu$m. The graphene parameters are $T=300$K, $\gamma_a=\gamma_b=0.1$meV and $\mu_a=\mu_b=0.5$eV. 
}\label{figura2}
\end{figure}
This equation shows the dependence of the eigenfrequencies with the gain--loss optical parameter $\varepsilon_i$. 
For $\varepsilon_i=0$, the real parts of the eigenfrequencies split by  $\Delta\,\omega=\omega_+ -\omega_- = 2 \frac{R^2}{R_{ab}^2} \frac{\omega_0}{(\varepsilon_1+\varepsilon_v)}  \varepsilon_v$ while their imaginary parts remain degenerated at the value $-\gamma_g/2$. From equation (\ref{sistema_ecuaciones_matricial_3}), we see that the eigenvector associated  to the upper branch verify $b_{a 1}=-b_{b 1}$ pointing out that the induced dipole moments on each of the cylinders move out of phase, \textit{i.e.},  the state corresponds to a trapped mode, while the eigenvector associated to the lower branch verify $b_{a 1}=b_{a 1}$ pointing out that this  mode corresponds to a radiating mode that can be excited,  for example, by an incident plane wave  \cite{CSZ}. We use the notation $+x$$-x$ and $+x$$+x$ to refer to the upper and lower branches, respectively.  
Since the terms inside the root have opposed signs between them, the splitting between upper ($+x$$-x$) and lower ($+x$$+x$) branches monotonically decreases as $\varepsilon_i$ is increased until reaching the value
\begin{eqnarray}\label{e_ic}
\varepsilon_{ep}= 2  \frac{R^2}{R_{ab}^2} \varepsilon_v, 
\end{eqnarray}
for which the eigenfrequencies coalesce at the exceptional point $\omega=-i\frac{\gamma_g}{2}+\frac{\omega_g}{\sqrt{\varepsilon_1+\varepsilon_v}}$. At the same time, the imaginary parts of both eigenfrequencies remain equal.  Beyond the exceptional point and for the gain--loss parameter $\varepsilon_i>\varepsilon_{ep}$, the imaginary parts of the eigenfrequencies bifurcate, while their real parts remain degenerated. In Figure \ref{figura2} we  illustrate in the complex plane the above described trajectories for the upper and lower branches (\ref{omega}) as parametric functions of the gain--loss parameter $\varepsilon_i$. We have taken  the straight segment $ z=0,\,\mbox{with}\,\Re\,z>0$ as the cut line for the square root function in Eq. (\ref{omega}) so that $\sqrt{-w^2}=i w$ for $w$ real and positive.

We now consider the case of polarization along the $y$ axis, in which   each of the induced dipole moments are along $\pm y$ direction, \textit{i.e.}, the local magnetic field associated to each of the  cylinders has a dependence $\approx \cos \phi_j$ ($j=a,\,b$). As a consequence, the amplitudes $b_{j 1}=-b_{j -1}$ ($j=a,\,b$). In this case, the matrix equation for these amplitudes is
\begin{equation}\label{sistema_ecuaciones_matricial_4}
\left[
\begin{array}{llll}
1 & - 2\,s_a\,H'_1(z)  \\
- 2\,s_b\,H'_1(z) & 1 \\
\end{array} \right] 
\left( 
\begin{array}{lll}
b_{a 1}\\
b_{b 1}\\
\end{array} \right)=
\left( 
\begin{array}{lll}
0\\
0\\
\end{array} \right),
\end{equation}
where we have used $2 H'_1(z)/z=H_0(z)-H_2(z)$, and $H'_1(z)$ is the derivative of $H_1(z)$. Following the same steps as in the case of $x$ polarization, the matrix equation (\ref{sistema_ecuaciones_matricial_4}) takes the form 
\begin{equation}\label{sistema_ecuaciones_matricial_5}
\left[
\begin{array}{llll}
\omega-\omega_a &  A_a/R_{ab}^2  \\
 A_b/R_{ab}^2 & \omega-\omega_b \\
\end{array} \right] 
\left( 
\begin{array}{lll}
b_{a 1}\\
b_{b 1}\\
\end{array} \right)=
\left( 
\begin{array}{lll}
0\\
0\\
\end{array} \right),
\end{equation}
    where $A_j$ are given by Eq. (\ref{A}). Note that the matrix in Eq. (\ref{sistema_ecuaciones_matricial_5}) differs from that in Eq. (\ref{sistema_ecuaciones_matricial_3}) only in the signs of the non diagonal terms. Therefore, the eigenfrequencies corresponding to induced dipole moments along $y$ direction are formally given by Eq. (\ref{omega}). Unlike the $x$ polarization case, for which the upper frequency branch for $\varepsilon_i=0$ corresponds to a trapped  mode, From Eq. (\ref{sistema_ecuaciones_matricial_5}) and taking $\varepsilon_i=0$, we see that the eigenvector associated to the upper branch verify $b_{a1}=b_{b1}$, \textit{i.e.}, it corresponds to a  radiating mode. Conversely, it is straightforward to verify that the lower frequency branch corresponds to a trapped mode for which $b_{a1}=-b_{b1}$.  It is worth noting that in   this case, $\pm y$ oscillations, it is convenient to take the cut of the complex square root function $\sqrt{z}$ as  the straight line $z=i w$ ($w$ real and positive) so that $\sqrt{-w^2}=-i w$. In this way, beyond the exceptional point the upper branch $+y$$+y$ moves away from the real axis  whereas the lower branch $+y$$-y$ reaches the real axis, as shown in Figure \ref{figura2}b.

\section{Results}  \label{resultados}
 
We consider a system of two  dielectric cylinders with permittivities $\varepsilon_a=2.13+i \varepsilon_i$ ($\varepsilon_i>0$), $\varepsilon_b=2.13-i\varepsilon_i$ for lossy and gain cylinders, respectively.  The radii $R_a=R_b=R=0.03\mu$m and both cylinders are coated with a graphene monolayer and  immersed in vacuum ($\varepsilon_v=1$).  The graphene parameters are: temperature $T=300$K, chemical potentials $\mu_1=\mu_2=0.5$eV and the carriers scattering rates $\gamma_1=\gamma_2=0.1$meV. The positions of the cylinders are $\mathbf{r}^a=-0.05\mu$m$\hat{x}$, $\mathbf{r}^b=0.05\mu$m$\hat{x}$ (center to center distance $R_{ab}=0.1\mu$m) and the gap between them is $\Delta=0.04\mu$m. 

\begin{figure}[htbp!]
\centering
\resizebox{0.7\textwidth}{!}
{\includegraphics{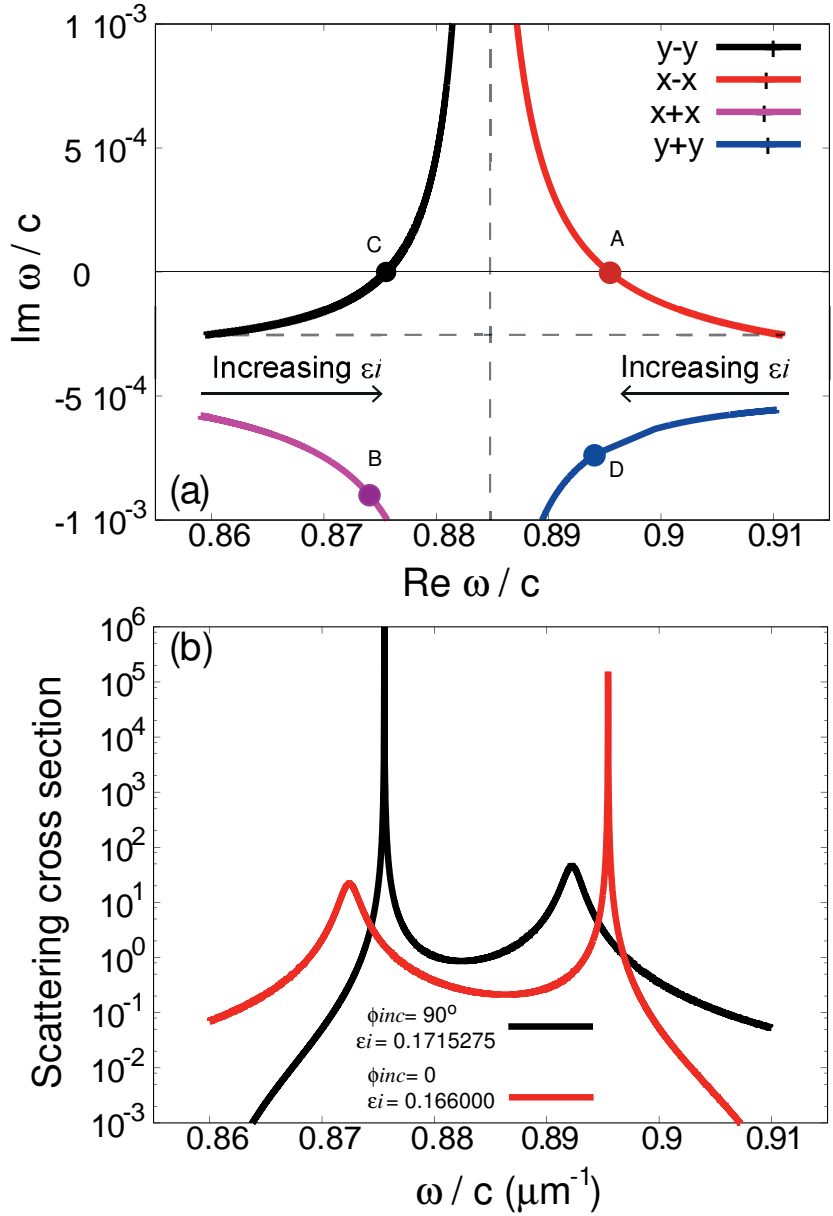}}
\caption{\label{fig:epsart} (a) Trajectory of the four branches as a parametric function of the gain--loss parameter $\varepsilon_i$. The lasing threshold is achieved for branches $+x$$-x$ and $+y$$-y$ at points A and C, respectively. The gain--loss parameter values are $\varepsilon_i=0.166$, for point A, and   $\varepsilon_i=0.1715275$, for point C. Points B and D correspond to states on the $+x$$+x$ and $+y$$+y$ branches for $\varepsilon_i=0.166$ and $\varepsilon_i=0.1715275$, respectively. Dashed lines correspond to QA branches plotted in Figure \ref{figura2}. (b) Scattering cross section for $p$--polarized incident waves for illumination direction along ($\phi_{inc}=90^\circ$) and  perpendicular ($\phi_{inc}=0$) to the axis joining the cylinder centers. The permittivity of the cylinders are $\varepsilon_a=2.13+i \varepsilon_i$, $\varepsilon_b=2.13-i \varepsilon_i$, the radius $R_a=R_b=R=0.03\mu$m and the gap $\Delta=0.04\mu$m. The graphene parameters are $T=300$K, $\gamma_a=\gamma_b=0.1$meV and $\mu_a=\mu_b=0.5$eV. 
}\label{figura3}
\end{figure}

Figure \ref{figura3}a shows the trajectory of the eigenfrequencies in the complex plane as a parametric function of the gain--loss parameter $\varepsilon_i$ calculated by solving the full retarded dispersion  equation (FR). To solve this equation, we use a Newton--Raphson method adapted to treat complex variable.  
Four branches are observed, two of them ($+x$$+x$ and $+x$$-x$) corresponds to both cylinders polarized along the $x$ axis (dipole moments along the $x$ axis) while the other two branches ($+y$$-y$ and $+y$$+y$) corresponds to the case in which both cylinders are polarized in the $y$ axis (dipole moments along $y$ axis). We are  using the notation of  the asymptotic case when $\varepsilon_i=0$ for naming the dimer surface plasmons, so that $+y$$+y$ ($+x$$+x$) configuration corresponds to the two dipole moments moving in phase on the $y$ axis ($x$ axis), and the $+y$$-y$ ($+x$$-x$)   configuration corresponds to the dipole moments oscillating in opposite phase on the $y$ axis ($x$ axis).  For instance, the branch $+x$$+x$ start at frequency $\omega/c=0.8589226-i0.5789274\,10^{-3}\mu$m$^{-1}$ for $\varepsilon_i=0$, where both dipole moments are in phase, and it moves to the right side leaving away from the real axis. Moreover, the  real part of this trajectory approaches asymptotically to the value $\omega/c=0.8868\mu$m$^{-1}$ corresponding to the real part of the eigenfrequency of a single graphene cylinder \cite{CRD}. On the contrary, the branch $+x$$-x$ start at $\omega/c=0.9107941-i0.253260710\,10^{-3}\mu$m$^{-1}$ for $\varepsilon_i=0$, with both dipole moments in opposite phase, and it approaches to the real axis where the lasing threshold at $\omega_{crit}/c \approx 0.8954615$m$^{-1}$ is reached for $\varepsilon_i=\varepsilon_{crit}=0.166$ (point A in Figure \ref{figura3}a).  

On the other hand, Figure \ref{figura3}a also shows the trajectory of the branch corresponding to the polarization along the  $y$ axis.  The branch $+y$$-y$ start at frequency $\omega/c=0.8594583-i0.25349871\,10^{-3}\mu$m$^{-1}$ for $\varepsilon_i=0$, where both dipole moments are in opposed  phase, and it moves toward the right side, reaching the real axis at point C where the critical  eigenfrequency $\omega_{crit}/c=0.8755407\mu$m$^{-1}$ for a critical value of the gain--loss parameter  $\varepsilon_{crit}=0.1715275$. On the other hand, the $+y$$+y$ branch start at frequency $\omega/c=0.9102954-i0.5560595\,10^{-3}\mu$m$^{-1}$ for $\varepsilon_i=0$ and it moves to the left side leaving away from the real axis. As in the $x$ polarized case, both branches are repelled in the direction of the imaginary axis allowing the $+y$$-y$ branch to achieve the lasing threshold at point C. 

In order to understand the gain--loss compensation near the critical points at which the eigenfrequencies are almost real, in Figure \ref{figura3}b we plot the scattering cross sections 
for a plane wave impinging at an angle $\phi=0$ (electric field along the $x$ axis)  and $\phi=90^\circ$ (electric field along the $y$ axis). The corresponding gain--loss parameter is $\varepsilon_i=0.166$ for $\phi=0$, and $\varepsilon_i=0.1715$  
for $\phi=90^\circ$, \textit{i.e.}, it values are near to the critical values at which lasing conditions are achieved. In the $\phi=0$ case, we observe that the scattering cross section is enhanced at frequency near $\omega/c \approx 0.895$m$^{-1}$ that agree well with the lasing frequency for the $+x$$-x$ branch calculated by solving the eigenmode problem (point A in Figure \ref{figura3}a). Moreover, we observe another peak (less intense) near $\omega/c=0.873\mu$m$^{-1}$, a value that falls near the real part of the eigenfrequency of an state with $\varepsilon_i=0.166$ but corresponding to the $+x$$+x$ branch (point B in Figure \ref{figura3}a). A similar behavior presents the scattering curve for $\phi=90^\circ$ and  $\varepsilon_i=0.1715$. From Figure \ref{figura3}b we observe a very sharp peak at frequency  that coincides with the lasing frequency $\omega_{crit}/c \approx 0.8755407\mu$m$^{-1}$ calculated by solving the eigenmode problem (point C in Figure \ref{figura3}a).  Moreover, a second peak is observed at  a frequency $\approx 0.894\mu$m$^{-1}$ associated to the excitation of the state on the  $+y$$+y$ branch for $\varepsilon_i=0.1715$ (point D in Figure \ref{figura3}a).

\begin{figure}[htbp!]
\centering
\resizebox{0.7\textwidth}{!}
{\includegraphics{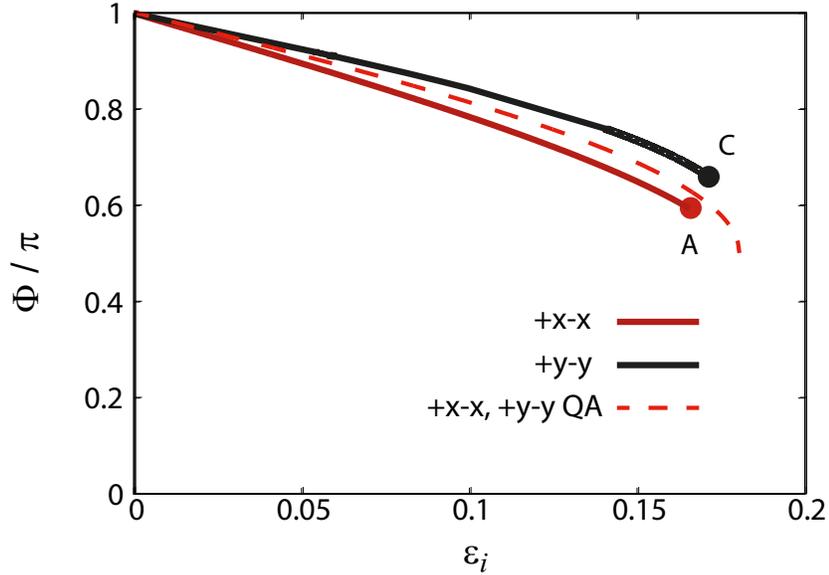}}
\caption{\label{fig:epsart} Phase difference  between the induced dipoles on each cylinders as a function of the gain--loss parameter $\varepsilon_i$ for the $+x$$-x$ and $+y$$-y$ branches. The calculations have been carried out by applying  the FR dispersion method 
(continuous line) and the QA method (dashed line).  Constitutive and geometrical parameters are the same as in Figure \ref{figura3}.
}\label{figura4}
\end{figure}

The question that arises from the above results is how eigenmodes on the $+x$$-x$ and $+y$$-y$ branches, which  correspond to trapped modes for $\varepsilon_i=0$, can be excited with a plane wave by varying the gain--loss parameter $\varepsilon_i$. Furthermore, these branches reach their lasing threshold for a critical value of the  gain--loss parameter. To find a response, we have calculated the eigenvectors, which contains all the field amplitudes of the eigenmodes. In particular, we have verified that coefficients with $|m|\not=1$ are orders of magnitude less than those corresponding to $m=\pm 1$, suggesting that the dimer plasmons can be considered, as a good approximation,  as a superposition of single plasmon with $m=\pm 1$ linked to each cylinder. 
As a consequence, to gain further insight into the underlying physics 
of these branches excitations, 
we applied the QA as follows. 
Without loss of generality, we consider the case for which the induced dipole moments on cylinders are in $\pm x$ direction. 
By replacing Eq. (\ref{omega}) into Eq. (\ref{sistema_ecuaciones_matricial_3})  and using Eq. (\ref{omega_1}), we find the following relation between the dimer amplitudes 
\begin{eqnarray}\label{amplitudes}
b_{a 1}= \mp \frac{\varepsilon_v (R/R_{ab})^2}{\sqrt{-\frac{\varepsilon_i^2}{4}+\varepsilon_v^2 (R/R_{ab})^4} + i \frac{\varepsilon_i}{2}} \, b_{b 1}, 
\end{eqnarray}
where the $-$ and $+$ signs correspond to the $+x$$-x$ and $+x$$+x$ branches, respectively.  From this equation we see that the modulus of the $b_{a 1}$ and $b_{b 1}$ amplitudes are equal providing that the gain--loss parameter be less than $\varepsilon_{ep}$. Taking into account the fact that $b_{a -1}=b_{a 1}$ and $b_{b -1}=b_{b 1}$ and using Eq. (\ref{campo_scattereado}), we can write the field scattered by the cylinders as
\begin{equation}\label{campo_scattereado_dipolar}
H^{(v)}_s(\mathbf{r}) = 2 i b_{a 1} H_1(k_v r_a) \sin(\phi_a) + 2 i b_{b 1} H_1(k_v r_b) \sin(\phi_b).
\end{equation}
Comparing this expression with that corresponding to a single dipole moment $p$ along the $x$ axis and centered at  the origin ($H \approx p H_1(k_v r) \sin(\phi)$), we deduce that Eq. (\ref{campo_scattereado_dipolar}) corresponds to a superposition of two fields, one of them due to a dipole moment of amplitude $p_a \approx b_{a 1}$ centered at the  cylinder $a$ and other due to a dipole moment of amplitude $p_b \approx b_{b 1}$ centered at the cylinder $b$. Since $b_{a 1}=b_{b 1}$ regardless of $\varepsilon_i$ ($\varepsilon_i<\varepsilon_{ep}$), the induced dipole moments $p_a=p_b$ and, as a consequence, both emitted fields for each of the dipoles have the same intensity. In addition, from Eq. (\ref{amplitudes}) we see that the phase difference $\Phi$ between $b_{a 1}$ and $b_{b 1}$ amplitudes, and thus between $p_a$ and $p_b$,  is shifted from $\pi$ to  $\pi/2$  when $\varepsilon_i$ increases from $0$ up to above  the  value $\varepsilon_{ep}$. 
This fact implies that the eigenmodes on the $+x$$-x$ trajectory pass from  trapped  to bright by increasing the gain--loss parameter.  %
Our calculation confirm this expectation, as can be seen in Figure \ref{figura4} where we show plots of the phase $\Phi$  
as a function of the gain--loss parameter $\varepsilon_i$ by applying the FR dispersion rigorous method (continuous line) and using Eq. (\ref{amplitudes})  (dashed line). 
From this Figure we see that the curve calculated with the QA agree well with that calculated by using the FR dispersion method. 
In particular, we see that the phase $\Phi=\pi$ for  $\varepsilon_i=0$, which means that the eigenmodes are dark states, and monotonically decreases for until reach the lasing modes A (for $+x$$-x$ brach) and B (for $+y$$-y$ branch).


It is worth noting  some similarities and differences between the eigenmode branches  calculation by using the FR dispersion equation  and those calculated by using the QA. On the one hand, the $+x$$+x$ and $+x$$-x$  branches in Figure \ref{figura3}a are repelled in the direction of the imaginary axis, as predicted by the QA for a PT--symmetric system (Figure \ref{figura2}), a feature that allows the $+x$$-x$ branch  to achieve the lasing threshold at point A. Furthermore, the critical value (\ref{e_ic}) calculated by using the QA results $\varepsilon_{ep}=0.18$, a value that agree well with the gain--loss parameter for which the lasing threshold is achieved in Figure \ref{figura3}a. Moreover, the  phase difference between the induced dipole moments on each cylinders calculated  by FR and QA methods matches quite well. On the other hand, $+x$$+x$ and $+x$$-x$ branches in Figure \ref{figura3}a do not start at points with the same value of their  imaginary parts  as occur for the branches in Figure \ref{figura2} and, as a consequence,  these trajectories does not coalesce at an exceptional point as in Figure \ref{figura2}. This is true because the lack of a radiation losses term in the QA, \textit{i.e.}, the radiation losses would prevent the system from having all the properties of a full loss compensated PT--symmetric structure, shown in Figure \ref{figura2}, such as the existence of an exceptional point.   

\begin{figure}[htbp!]
\centering
\resizebox{0.95\textwidth}{!}
{\includegraphics{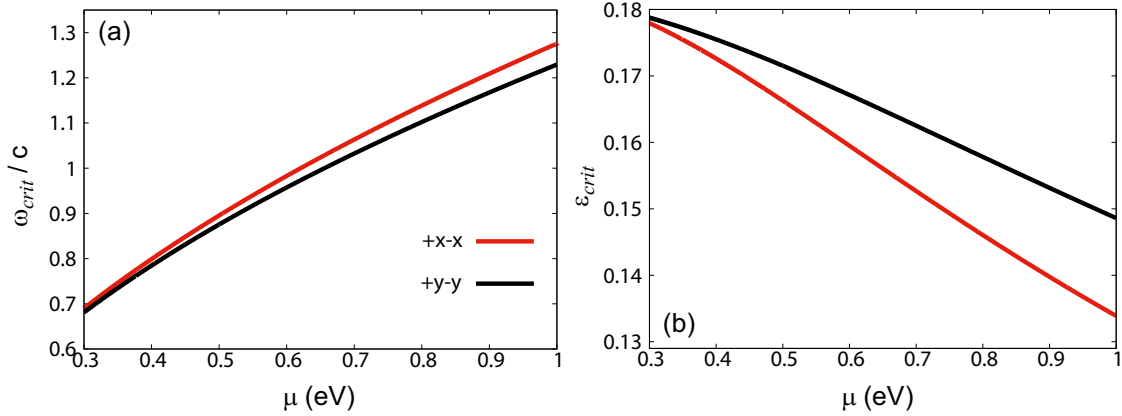}}
\caption{\label{fig:epsart} Lasing frequency (a) and critical gain--loss parameter (b) as functions of the chemical potential $\mu$ ($\mu_a=\mu_b=\mu$) for branches $+x$$-x$ and $+y$$-y$.  The calculations have been carried out by applying  the FR dispersion method. Geometrical and other constitutive parameters are the same as in Figure \ref{figura3}.
}\label{figura5}
\end{figure}

In order to study the behavior with the chemical potential on graphene, we set $\mu_a=\mu_b=\mu$ and vary the values of  $\mu$. In Figure \ref{figura5} we have plotted the lasing frequency $\omega_{crit}$ and the corresponding gain--loss parameter $\varepsilon_{crit}$ as functions of $\mu$, calculated with the rigorous FR method. From Figure \ref{figura5}a we can see that the lasing frequency for both $+x$$-x$ and $+y$$-y$ branches are increasing functions of $\mu$. This fact can be understood by taking into account that the real part of the eigenfrequency corresponding to a single graphene cylinder, which falls between the lasing frequencies for $+y$$-y$ and $+x$$-x$ branches, is proportional to $\sqrt{\mu}$ (see Eq. (\ref{omega_R})). On the other hand, from Figure \ref{figura5}b we see that the critical value of the gain--loss parameter for which the  lasing condition is achieved is a decreasing function of the chemical potential $\mu$. This behaviour has a similarity with that presented by a single graphene cylinder for which has been demonstrated that the gain level to achieve the lasing condition decreases with the chemical potential value \cite{Leila}.

\section{Conclusions} \label{conclusiones}

In conclusion, we have analytically studied  the scattering and the eigenmode problems for a dimer composed of two graphene coated dielectric cylinders, one of them with  loss  and the other with the same level of gain. We have demonstrated the existence of two branches, corresponding to trapped modes when $\varepsilon_i=0$, that reach the lasing conditions for suitable values of the  gain--loss parameter. While the phase difference between the induced dipole moments on individual cylinders changes from $\pi$ to a value near $\pi/2$ when the gain--loss parameter is incremented, a fact that provides the mode transformation from trapped to radiating modes, the modulus of each individual dipole moments maintains equals in between. 

Other mechanisms to transform a trapped mode into a resonant observable which can be excited by a plane wave have been reported in other works. 
All these methods are based on the introduction of a small asymmetry with respect to the center of the dimer by producing   dissimilar  dipole moment modulus on each of the cylinders. 
Interestingly, here, we found that in the transformation from trapped to radiating  eigenmodes on both  $+x$$-x$ and $+y$$-y$  branches,  the modulus of the individual dipole moments does not change, while  it is changed the phase between them. A distinction between these two kind of mechanism,  the phase variation and modulus variation mechanisms, to transform a trapped mode into a resonant observable  has not been  reported before to our knowledge. We believe that our results will be usefull for a deeper  understanding of the PT--symmetric LSP characteristics and that the LSP effects we have demonstrated opens up possibilities for practical applications involving sub--wavelength laser structures.       

\section*{Funding}
Consejo Nacional de Investigaciones Cientficas y T\'ecnicas (CONICET)

\section*{Acknowledgments}
The authors acknowledge the financial support of Consejo Nacional de Investigaciones Cientficas y T\'ecnicas (CONICET)

\section*{Disclosures}

The authors declare no conflicts of interest.

\appendix

\section{Graphene conductivity} \label{grafeno}
\setcounter{equation}{0}
\renewcommand{\theequation}{A{\arabic{equation}}}
We consider the  graphene layer as an infinitesimally thin, local and isotropic two--sided layer with frequency--dependent surface conductivity $\sigma(\omega)$ given by the Kubo formula \cite{falko}, which can be read as  $\sigma= \sigma^{intra}+\sigma^{inter}$, with the intraband and interband contributions being
\begin{equation} \label{intra}
\sigma^{intra}(\omega)= \frac{2i e^2 k_B T}{\pi \hbar^2 (\omega+i\gamma_g)} \mbox{ln}\left[2 \mbox{cosh}(\mu_g/2 k_B T)\right],
\end{equation}  
\begin{eqnarray} \label{inter}
\sigma^{inter}(\omega)= \frac{e^2}{\hbar} \Bigg\{   \frac{1}{2}+\frac{1}{\pi}\mbox{arctan}\left[(\hbar \omega-2\mu_g)/2k_BT\right]-\nonumber\\
\frac{i}{2\pi}\mbox{ln}\left[\frac{(\hbar \omega+2\mu_g)^2}{(\hbar \omega-2\mu_g)^2+(2k_BT)^2}\right] \Bigg\},
\end{eqnarray}  
where $\mu_g$ is the chemical potential (controlled with the help of a gate voltage), $\gamma_g$ the carriers scattering rate, $e$ the electron charge, $k_B$ the Boltzmann constant and $\hbar$ the reduced Planck constant.

\section{Developing of the right hand side in equation (\ref{dispersion_2})} \label{desarrollo}
\setcounter{equation}{0}
\renewcommand{\theequation}{B{\arabic{equation}}}
The right side of Eq. (\ref{dispersion_2}) can be calculated by using Eq. (\ref{A}) as follows.  
\begin{eqnarray}\label{AB}
\frac{A_a A_b}{R_{ab}^4} = \frac{a^4}{R_{ab}^4} \frac{[\varepsilon_1+i\varepsilon_i-\varepsilon_v+g(\omega_a)]}{g'(\omega_a)} \frac{[\varepsilon_1-i\varepsilon_i-\varepsilon_v+g(\omega_b)]}{g'(\omega_b)}, 
\end{eqnarray}
where 
\begin{eqnarray}\label{gj}
g(\omega_j)=-\frac{\omega_g^2}{\omega_j^2+i\gamma_g \omega_j}, 
\end{eqnarray}
and
\begin{eqnarray}\label{dgj}
g'(\omega_j)=\frac{\omega_g^2 (2 \omega_j+i \gamma_g)}{(\omega_j^2+i\gamma_g \omega_j)^2}.
\end{eqnarray}
By taking into account that $\gamma_g<<|\omega_j|$, we can write this equations as,
\begin{eqnarray}\label{gj_approx}
g(\omega_j)=-\frac{\omega_g^2}{\omega_j^2}\frac{1}{1+i\gamma_g/\omega_j}\approx -\frac{\omega_g^2}{\omega_j^2} \left(1-i\frac{\gamma_g}{\omega_j}+\frac{\gamma_g^2}{\omega_j^2}+...\right), 
\end{eqnarray}
and 
\begin{eqnarray}\label{dgj_approx}
g'(\omega_j) \approx \frac{ 2 \omega_g^2}{\omega_j^3} \left(1-i\frac{\gamma_g}{\omega_j}+\frac{\gamma_g^2}{\omega_j^2}+...\right)-\frac{\omega_g^2}{\omega_j^2} \left(i\frac{\gamma_g}{\omega_j^2}-2 \frac{\gamma_g^2}{\omega_j^3}+... \right) =\nonumber\\
 \frac{2 \omega_g^2}{\omega_j^3} \left( 1 - \frac{3}{2}i\frac{\gamma_g}{\omega_j}+2\frac{ \gamma_g^2}{\omega_j^2}+... \right). 
\end{eqnarray}
Considering the lowest order in $\Gamma_{sup}/\omega_0$ and $\Gamma_{cover}/\omega_0$, functions $g(\omega_j)$ and $g'(\omega_j)$ are written as
\begin{eqnarray}\label{gj_0}
g(\omega_j)\approx-\frac{\omega_g^2}{\omega_0^2} \left[1 \pm \frac{i}{\omega_0}2\Gamma_{core}\right],
\end{eqnarray}
and
\begin{eqnarray}\label{dgj_0}
g'(\omega_j) \approx 
 \frac{2 \omega_g^2}{\omega_0^3} \left[1 \pm  i\frac{3}{\omega_0}\Gamma_{core}\right].
\end{eqnarray}
Signs $+$ and $-$ correspond to $j=a$ and $j=b$, respectively.
%
In this approximation, Eq. (\ref{AB}) reduces to
\begin{eqnarray}\label{lado_derecho}
\frac{A_a A_b}{R_{ab}^4} = 
\frac{R^4}{R_{ab}^4} \frac{\omega_g^2}{(\varepsilon_1+\varepsilon_v)} \varepsilon_v^2. 
\end{eqnarray}

\end{document}